# A description of odd mass W-isotopes in the Interacting Boson-Fermion Model

S. Abu Musleh[1] and O. Scholten[2]

[1] *National Center Of Research Palstine - Gaza*

[2] *Kernfysisch Versneller Instituut, University of Groningen,*

*9747 AA, Groningen, The Netherlands*

(Dated: November 14, 2018)

## Abstract

The negative and positive parity low-spin states of the even-odd Tungsten isotopes, $^{183,185,187}W$ are studied in the frame work of the Interacting Boson-Fermion Approximation (IBFA) model. The fermion that is coupled to the system of bosons is taken to be in the negative parity $2f_{7/2}$, $2f_{5/2}$, $3p_{3/2}$, $3p_{1/2}$ and in the positive parity $1i_{13/2}$ single-particle orbits. The calculated energies of low-spin energy levels of the odd isotopes are found to agree well with the experimental data. Also B(E2) values and spectroscopic factors for single-neutron transfer are calculated and found to be in good agreement with experimental data.

## I. INTRODUCTION

In recent papers [1–3] detailed level schemes of $^{183}W$ and $^{187}W$ are presented including recent new data on spectroscopic factors for the single neutron stripping reactions as obtained from $(n,\gamma)$ and $(\vec{d},p)$ reactions. This has stimulated us to perform systematic calculations in the Interacting Boson-Fermion Approximation (IBFA) model [4] for the isotopes $^{183}W$, $^{185}W$, and $^{187}W$.

In the Interacting Boson Model (IBM) [5] the low-lying states in even-even nuclei are described in terms of the relevant effective degrees of freedom which are nucleon pairs coupled to angular momentum $J=0$ and $J=2$, accounted for in terms of $s$ and $d$ bosons. This allows to incorporate the important aspects of the interaction, being the pairing force between like nuclei and a quadrupole interaction between neutrons and protons. For certain limiting choices for the model parameters, corresponding to dynamical symmetries, the spectrum corresponds to that of an (an)harmonic vibrator, an axial-symmetric rotor and that of a triaxial gamma-unstable rotor [4]. The existence of these dynamical symmetries has contributed greatly to the success of the model.

Relatively many experimental [6–11] studies have been devoted to the even-even nuclei in the mass region of the neutron rich Tungsten isotopes, in stark contrast with the little attention given to the odd-mass isotopes [12–16]. Woods-Saxon-potential based calculations [17] have suggested the prolate-oblate shape change takes place between $N=114$ and $N=116$, with similar calculations suggesting $N=118$ [8]. In earlier IBM studies [6, 10] the even mass Tungsten isotopes have been calculated and we will compare our parameters with these.

For this work we are particularly interested in the calculation of odd-mass nuclei in the IBFA model [18]. The addition of the degrees of freedom of a single particle to those of the $s$ and $d$ bosons allows for a more detailed test of the microscopic basis of the model since the structure of the Hamiltonian, and in particular that of the single-particle-transfer operator, strongly depends on the interpretation of bosons in terms of fermion pairs following semi-microscopic arguments [19–21]. So far the calculation of spectroscopic factors in the IBFA model has not received much attention. For certain limiting cases the IBFA Hamiltonian has dynamical (super) symmetries [22–26], corresponding to the spectra of certain odd-mass nuclei. The model has also been extended to describe odd-odd nuclei [27].

We investigate a series of odd-mass tungsten isotopes, $^{183,185,187}W$, which have 74 protons and 109-113 neutrons and can be described in the IBFA by the coupling of the degrees of freedom of a single neutron to the even-even cores. Because of the vicinity of the $N = 126$ shell closure the neutrons are considered as holes rather than particles. In the major shell (N = 82-126), there are five single-particle levels, four with negative parity, $2f_{7/2}$, $2f_{5/2}$, $3p_{3/2}$, $3p_{1/2}$ and one with positive parity $1i_{13/2}$.

We first discuss the calculation of the even-even cores in the IBM. In Section III a quick overview of the most important ingredients of the IBFA model are reviewed. The calculated excitation energies are presented in Section IV followed in Section V with E2 transition probabilities. The results for spectroscopic factors for single neutron transfer are given in Section VI. The conclusions are presented in Section VII.

## II. THE EVEN-EVEN CORE

The IBM [5] provides a unified description of collective nuclear states in terms of a system of interacting bosons. The IBM-1 (the version of the model where proton and neutron bosons are not distinguished) Hamiltonian which we use to describe the even-even nuclei has the standard form as given in [5]. In terms of s- and d-bosons the most general IBM-1 Hamiltonian can be expressed as

$$\begin{aligned} H_B &= \varepsilon n_d + \frac{1}{2}\kappa_1(L \cdot L) + \frac{1}{2}\kappa_Q(Q_B \cdot Q_B) - 5\sqrt{7}\kappa_3[(d^\dagger \tilde{d})^{(3)}(d^\dagger \tilde{d})^{(3)}]_0^{(0)} \\ &\quad +15\kappa_4[(d^\dagger \tilde{d})^{(4)}(d^\dagger \tilde{d})^{(4)}]_0^{(0)} , \end{aligned} \quad (1)$$

where

$$(\vec{L} \cdot \vec{L}) = -10\sqrt{3}[(d^\dagger \tilde{d})^{(1)}(d^\dagger \tilde{d})^{(1)}]_0^{(0)} , \quad (2)$$

and

$$Q_B = \{(s^\dagger \tilde{d} + d^\dagger s)^{(2)} + \frac{\chi}{\sqrt{5}}(d^\dagger \tilde{d})^{(2)}\} . \quad (3)$$

For the diagonalization of the Hamiltonian we used the computer code PHINT [32]. The values of the parameters in the Hamiltonian for the even mass W-isotopes are obtained from a fit to the experimental energies [33] of the three lowest bands (ground state, $\beta$, and $\gamma$-bands). The values of the interaction parameters in the IBM-1 Hamiltonian which gave the best fit to the experimental data are given in Table I. The trend of the present IBM-1 parameters across the $^{182-186}W$ isotopes is in excellent agreement with that of the

TABLE I: The IBM-1 parameters as used in our calculations. All parameters are in keV except $\chi$ which is dimensionless.

| Isotope | $N_\pi + N_\nu$ | $\varepsilon$ | $\kappa_Q$ | $\kappa_1$ | $\kappa_3$ | $\kappa_4$ | $\chi$ |
|---|---|---|---|---|---|---|---|
| $^{182}W$ | 13 | 6. | -39. | 18.85 | 6.4 | 14 | -1.90 |
| $^{184}W$ | 12 | 15. | -35. | 17.4 | 8.0 | 9.9 | -1.75 |
| $^{186}W$ | 11 | 17. | -31. | 17.1 | 9.5 | 9.5 | -1.71 |

parameters for the Er isotopes which have the same number of bosons and lie in the same mass region [29]. They also agree well with those of ref. [6]. Our result indicate a shape change in the W isotopes from O(6) to SU(3). The calculated energy levels are shown in Fig. ?? where they are compared with the experimental data. height

Further insight in the structure is obtained from the B(E2) values. The E2 transition operator can be written as [5]

$$T(E2) = e_b Q_B , \qquad (4)$$

where $Q_B$ has been defined in Eq. (3) and $e_b$ is the boson effective charge. Fitting the absolute B(E2) strengths for transitions within the ground-state band resulted in $e_b = 0.13$ eb for all isotopes. The obtained B(E2) values are compared with experimental data in Table II showing a very good agreement.

## III. THE INTERACTING BOSON-FERMION APPROXIMATION MODEL.

In the IBFA model [4], odd-A nuclei are described by the coupling of the degrees of freedom of the low-energy quasi-particle levels to a collective boson core. The total Hamiltonian can be written as the sum of three parts,

$$H = H_B + H_F + V_{BF} , \qquad (5)$$

where $H_B$ is the usual IBM-1 Hamiltonian [5] for the even-even core, $H_F$ is the fermion Hamiltonian containing only one-body terms, and $V_{BF}$ is the boson-fermion interaction that describes the interaction between the odd quasi-nucleon and the even-even core nucleus [21]. The one-body term can be written as

$$H_F = \sum_{jm} \varepsilon_j a^\dagger_{jm} a_{jm} , \qquad (6)$$

TABLE II: Calculated B(E2) values (in $e^2b^2$) for transitions in the $^{182,184,186}W$ isotopes are compared to experimental data [30] and a previous work by Duval and Barrett [6]

|  | This work | | | Duval & Barrett | | | Experiment | | |
|---|---|---|---|---|---|---|---|---|---|
|  | 182 | 184 | 186 | 182 | 184 | 186 | 182 | 184 | 186 |
| $2_1^+ \to 0_1^+$ | 0.843 | 0.723 | 0.517 | 0.84 | 0.67 | 0.53 | 0.84 | 0.72 | 0.51 |
| $2_2^+ \to 0_1^+$ | 0.018 | 0.0156 | 0.0104 | 0.022 | 0.033 | 0.031 | 0.0248 | 0.0252 | 0.03 |
| $2_3^+ \to 0_1^+$ | 0.012 | 0.010 | 0.0091 | 0.0057 | 0.0011 | 0.0018 | 0.0065 | 0.002 | 0.0016 |
| $2_2^+ \to 2_1^+$ | 0.058 | 0.078 | 0.098 | 0.056 | 0.0641 | 0.0644 | 0.065 | 0.05 | 0.064 |
| $2_3^+ \to 2_1^+$ | 0.0062 | 0.0028 | 0.0005 | 0.0022 | 0.0014 | 0.0024 | 0.0057 | | |
| $4_1^+ \to 2_1^+$ | 1.19 | 1.032 | 0.742 | 1.19 | 0.949 | 0.746 | 1.16 | 1.03 | 0.905 |
| $6_1^+ \to 4_1^+$ | 1.28 | 1.12 | 0.81 | | | | 1.28 | 1.14 | 1.17 |
| $8_1^+ \to 6_1^+$ | 1.33 | 1.15 | 0.81 | | | | 1.28 | 1.76 | 1.12 |
| $10_1^+ \to 8_1^+$ | 1.33 | 1.13 | 0.80 | | | | 1.04 | 1.80 | 0.9 |

where $\varepsilon_j$ denotes the quasi-particle energies and $a_{jm}^\dagger$ and $a_{jm}$ are the creation and annihilation operators for the quasi-particle in the eigenstate $|jm\rangle$. The boson-fermion interaction $V_{BF}$ is described in terms of three contributions; i) a monopole interaction which is characterized by the parameter $A_0$, ii) a quadruple interaction [5, 31] characterized by $\Gamma_0$, and iii) the exchange of a quasi particle with one of the two fermions forming a boson [5] characterized by $\Lambda_0$,

$$V_{BV} = \sum_j A_j[(d^\dagger \tilde{d})^{(0)}(a_j^\dagger \tilde{a}_j)^{(0)}] + \sum_{jj'} \Gamma_{jj'}[Q^{(2)}(a_j^\dagger \tilde{a}_j)^{(2)}]_0^{(0)} \\ + \sum_{jj'j''} \Lambda_{jj'}^{j''} : [(d^\dagger \tilde{a}_j)^{(j'')}(a_{j'}^\dagger \tilde{d}_j)^{(j'')}]_0^{(0)} :, \quad (7)$$

where

$$Q = \{(s^\dagger \tilde{d} + d^\dagger s)^{(2)} + \frac{\chi}{\sqrt{5}}(d^\dagger \tilde{d})^{(2)}\}, \quad (8)$$

$\tilde{a}_{jm} = (-1)^{j-m}a_{j-m}$, and :: denotes normal ordering whereby contributions that arise from commuting the operators are omitted. The first term in $V_{BF}$ is the monopole interaction which plays a minor role in the actual calculations. The dominant terms in Eq. (7) are the second and the third terms, which both arise from the microscopic neutron-proton quadrupole interaction. For this reason the structure of the quadrupole operator may be different from the one used in the calculation of E2 transition probabilities. The third term, the exchange force, represents the exchange of a quasi particle with one of the two fermions forming a boson. It should thus be regarded as an effective contribution to the interaction resulting from the Pauli principle at the microscopic level in conjunction with the quadrupole interaction between protons and neutrons [21, 34]. The remaining parameters in Eq. (7) can be related to the BCS occupation probabilities $v_j$ of the single-particle orbits,

$$\Gamma_{jj'} = \sqrt{5}\Gamma_0(u_j u_{j'} - v_j v_{j'})Q_{jj'},$$
$$\Lambda_{jj'}^{j''} = -\sqrt{5}\Lambda_0[(u_{j'}v_{j''} + v_{j'}u_{j''})Q_{j'j''}\beta_{j''j} + (u_{j'}v_{j''} + v_{j'}u_{j''})Q_{j'j}\beta_{j'j''}]/\sqrt{2j''+1}, \quad (9)$$

where $Q_{j'j''}$ are single particle matrix elements of the quadruple operator and

$$\beta_{jj'} = (u_j v_{j'} + v_j u_{j'}) Q_{jj'}, \quad (10)$$

are the structure coefficients of the $d$-boson deduced from microscopic considerations [21]. The BCS occupation probabilities and the quasi-particle energy of each single-particle orbital can in principle be obtained by solving the gap equations. In the present calculations we have taken the quasi-particle energies as well as the occupation probabilities as free parameters in addition to the strengths $\Lambda_0$, $\Gamma_0$ and $A_0$ to obtain the best fit to the excitation energies. The reason for taking this approach is that for these triaxial nuclei the microscopic structure of the bosons may be more complicated than what has been assumed for the microscopic structure of the boson-fermion interaction.

## IV.  EXCITATION ENERGIES

In the present study of the $^{183-187}W$ isotopes the parameters in the Hamiltonian were adjusted to obtain the best overall agreement with the measured excitation energies of positive and negative parity states. We have opted to keep the coupling strengths constant for the three isotopes, resulting in the Boson-Fermions parameters $A_0 = -0.26$, $\Gamma_0 = -0.1$, $\chi = -.213$, and $\Lambda_0 = 0.55$ in Eq. (5). For the calculations the computer program ODDA [35] in which the IBFA parameters are identified as $A_0 = BFM$, $\Gamma_0 = BFQ$ and $\Lambda_0 = BFE$. The single particle occupation probabilities and the quasiparticle energies were allowed to vary across the range of isotopes. As can be seen from Table III the parameters show a gradual change. For the calculation of excitation energies only the relative values of the quasi-particle energies are important for which reason we have normalized to lowest one to zero for each isotope.

TABLE III: Occupation probabilities and quasi-particle energies for the $2f_{7/2}$, $2f_{5/2}$, $3p_{3/2}$, $3p_{1/2}$ and the $1i_{13/2}$ single particle orbits as used in the calculation of the $^{183,185,187}W$ isotopes.

|  | $^{187}W$ | | $^{185}W$ | | $^{183}W$ | |
| --- | --- | --- | --- | --- | --- | --- |
| $j$ | $v_j^2$ | $\varepsilon_j$ | $v_j^2$ | $\varepsilon_j$ | $v_j^2$ | $\varepsilon_j$ |
| $2f_{7/2}$ | 0.01 | .83 | 0.02 | 0.74 | .025 | 1.11 |
| $2f_{5/2}$ | 0.68 | 0.0 | 0.83 | 0.01 | 0.62 | 0.49 |
| $3p_{3/2}$ | 0.7 | .62 | 0.9 | 0.14 | 0.97 | 0.31 |
| $3p_{1/2}$ | .005 | 1.6 | 0.01 | 0.24 | 0.15 | 0.0 |
| $1i_{13/2}$ | 0.3 | .38 | 0.4 | 0.0 | 0.23 | 0.5 |

In the choice of the model space we have limited ourselves to the single-particle levels that are expected to play a dominant role for these nuclei. Based on the single particle energies as given in [28] we have decided to include the $2f_{7/2}$ level and exclude the $1h_{9/2}$ orbit. With its much larger quasi-particle energy it is expected that the influence of the $1h_{9/2}$ orbit on the energy spectrum will be small.

The structure of the spectrum is strongly determined by the strength of the exchange force, $\Lambda_0$, which agrees well with the values obtained from previous calculations in the W-mass region [29]. The value of the other two parameters shows larger differences, which could be due to the influence of including the $2f_{7/2}$ orbital in the model space. It should be notes that the same strength and parametrization is used for the interaction of the positive and negative parity single particle orbits with the core.

The calculated excitation energies are compared with experiment for the odd mass $^{183-187}W$ isotopes in Fig. 1–?? for positive and negative parity states. In general a very good agreement is obtained for the spectra of all three isotopes up to rather large excitation energies although there are some differences in the band-staggering patterns.

For $^{183}W$ the negative parity levels are reproduced well in the calculation. For the positive parity states the $13/2^+$ band is predicted at about the same energy as the $11/2^+$ band, however this band is not seem in experiment. A very similar effect is also seen in $^{185}W$ and $^{187}W$ which may be due to a too small value for the strength of the exchange force. Since there is only a single positive parity orbit coupled to the bosons this space is probably too small to create a realistic amount of collectivity. In the calculations this could be accounted for by increasing the strength of the boson-fermion interaction which lies however outside the scope of the present calculations. The difficulty in interpreting the higher lying bands is that in the data there are relatively low-lying levels with uncertain spin and parity assignments. The most prominent example is the level at 510 keV in $^{187}W$.

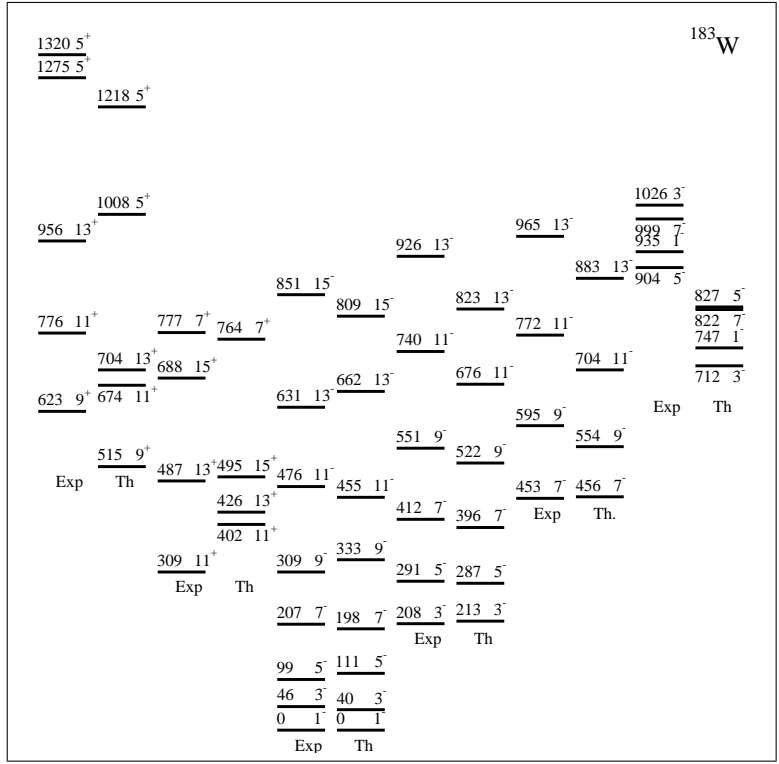

FIG. 1: Calculated energies for $^{183}W$ are compared to data. For each level the excitation energy in keV is given as well as the spin ($\times$ 2) and parity.

## V. ELECTROMAGNETIC TRANSITION PROBABILITIES

Electromagnetic transition rates form a good measure of the collective structure of the model wave functions. In general, the electromagnetic transition operators can be written as the sum of two terms, the first of which acts only on the boson part of the wave function

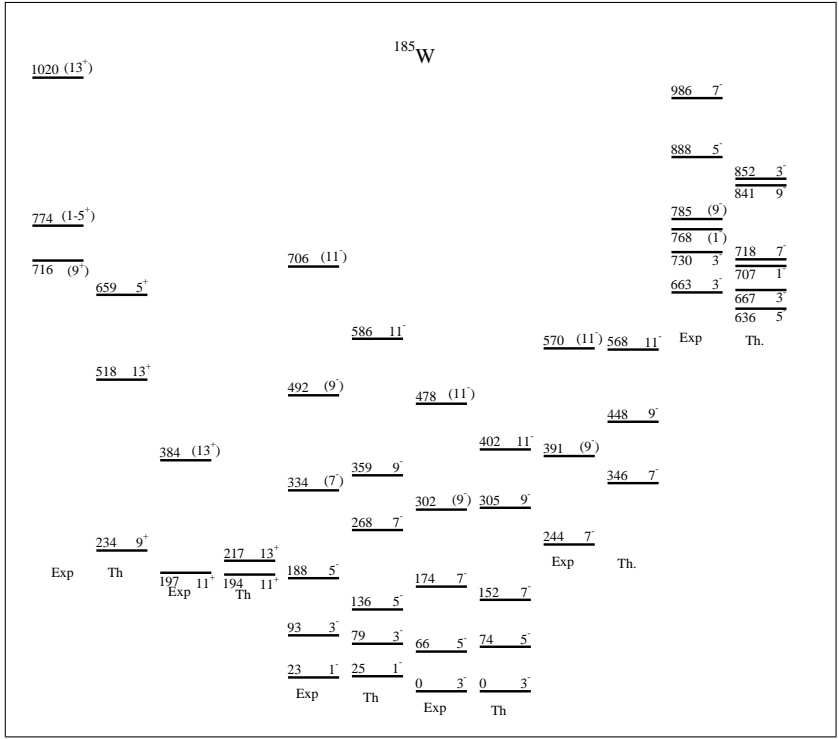

FIG. 2: Same as in Fig. 1 but for $^{185}W$.

and second only on the fermion part,

$$T^{(E2)} = e_b Q_B^{(2)} + e_\nu \sum_{jj''} Q_{jj'} (a_j \tilde{a}_j)^{(2)} ,  \qquad (11)$$

where $Q_B$ has been defined in Eq. (3), $Q_{j'j''}$ are single particle matrix elements of the quadruple operator which have also been used in the Hamiltonian Eq. (7), and $e_b$ and $e_\nu$ are the boson and fermion effective charges respectively. In the actual calculations the computer program PBEM [32] has been used. The boson effective charge has been determined earlier from the calculation of the even-even cores, $e_b = 0.13$ eb. The fermion effective charge is set equal to the boson effective charge, $e_\nu = 0.13$ eb, leaving no free parameters to adjust to the data. Experimental B(E2) values for transitions between negative parity states are compared to the results of our calculations in Table IV. In general a good agreement with the data is obtained however the data are relatively scarce.

TABLE IV: Calculated and experimental B(E2) values in units of $e^2b^2$.

| Isotope | Transition | Exp. | Theory |
|---|---|---|---|
| $^{183}W$ | $\frac{5}{2}_1^- \to \frac{1}{2}_1^-$ | 0.685 | 0.796 |
|  | $\frac{3}{2}_1^- \to \frac{1}{2}_1^-$ | 0.938 | 0.806 |
|  | $\frac{5}{2}_1^- \to \frac{1}{2}_1^-$ | 0.283 | 0.22 |
|  | $\frac{5}{2}_2^- \to \frac{3}{2}_2^-$ | 1.29 | 1.4 |
|  | $\frac{5}{2}_2^- \to \frac{5}{2}_1^-$ | 0.027 | 0.003 |
|  | $\frac{5}{2}_3^- \to \frac{1}{2}_1^-$ | 0.00246 | 0.0041 |
|  | $\frac{7}{2}_1^- \to \frac{5}{2}_1^-$ |  | 0.117 |
|  | $\frac{13}{2}_1^- \to \frac{9}{2}_1^-$ | 1.052 | 1.2 |
| $^{185}W$ | $\frac{7}{2}_1^- \to \frac{3}{2}_1^-$ | >0.11 | 0.5827 |
|  | $\frac{7}{2}_2^- \to \frac{3}{2}_1^-$ | 0.00151 | 0.0144 |
|  | $\frac{3}{2}_1^- \to \frac{1}{2}_1^-$ |  | 0.0632 |
|  | $\frac{3}{2}_2^- \to \frac{1}{2}_1^-$ |  | 0.724 |
| $^{187}W$ | $\frac{7}{2}_2^- \to \frac{5}{2}_1^-$ | 0.0059 | 0.0031 |
|  | $\frac{7}{2}_2^- \to \frac{7}{2}_1^-$ | 0.00889 | 0.001 |
|  | $\frac{3}{2}_1^- \to \frac{1}{2}_1^-$ |  | 0.0162 |

## VI. SPECTROSCOPIC FACTORS

Spectroscopic factors test the quasi-particle structure of the model wave functions and thus supply complementary information to that obtained from E2 transitions. The structure of the operator for single particle transfer in the model is directly based on the microscopic interpretation of the bosons as fermion pairs. In one particle transfer the (generalized) seniority of the state may increase or decrease with one unit. Since the quasi-particle operator $a_j^\dagger$ increases the generalized seniority of the state by one unit and $s^\dagger d a_j^\dagger$ decreases it [20], the general structure of the single particle transfer operator can be written as

$$c_j^\dagger = \left\{ u_j a_j^\dagger - \sqrt{\frac{10}{2j+1}} \frac{\sqrt{(N_\pi)}}{N} v_j \sum_{j'} \frac{\beta_{j',j}}{\sqrt{\mathcal{N}_\beta}} \left[ s^\dagger \tilde{d} a_{j'}^\dagger \right]^j \right\} / K_j \qquad (12)$$

TABLE V: Spectroscopic factors for one neutron transfer from the ground state of $^{186}W$ to various excited states in $^{187}W$ are compared with the data [2].

| $^{187}W$ | Exp. | | Theo. | |
|---|---|---|---|---|
| $j^\pi$ | E | $S_j \times 100$ | E | $S_j \times 100$ |
| $\frac{1}{2}^-_1$ | 0.1459 | .6 | 0.154 | 0.1 |
| $\frac{1}{2}^-_2$ | 0.762 | .1 | 0.755 | 0.0 |
| $\frac{3}{2}^-_1$ | 0.00 | 1.2 | 0.00 | .05 |
| $\frac{3}{2}^-_2$ | 0.204 | 9.8 | 0.214 | 3.0 |
| $\frac{3}{2}^-_3$ | 0.816 | 10. | 0.768 | 5.0 |
| $\frac{5}{2}^-_1$ | 0.077 | 14 | 0.074 | 19. |
| $\frac{5}{2}^-_2$ | 0.303 | 1.4 | 0.295 | 9.0 |
| $\frac{7}{2}^-_1$ | 0.2 | – | 0.172 | .04 |
| $\frac{7}{2}^-_2$ | 0.350 | 8.6 | 0.384 | .04 |
| $\frac{7}{2}^-_3$ | 0.432 | 2.8 | 0.399 | .05 |
| $\frac{13}{2}^+_1$ | 0.598 | 5.7 | 0.399 | 17. |

with $\mathcal{N}_\beta = \sum_{j,j'} \beta^2_{j',j}$ where the normalization constant $K_j$ is chosen such that

$$\sum_j \langle |c_j^\dagger| \rangle^2 = (2j+1)u_j^2 \tag{13}$$

The calculation of spectroscopic factors is free from adjustable parameters since the occupation probabilities have been determined from the calculation of excitation energies. For single-neutron stripping to $^{183}W$ an extensive list of spectroscopic factors was published recently [1]. Also for $^{187}W$ there are recent data available [2].

In Table V the results of our calculations are compared with these data for $^{187}W$. The rather good agreement indicates that our calculations give reliable wave functions for the low lying states. The results also show that the used occupation probability for the $f_{7/2}$ orbit, see Table III, was too low while that of the $f_{7/2}$ orbit is on the high side. The occupancies of the other orbits give a fair agreement with the data.

The data for $^{183}W$ are compared to the calculation in Table VI. The spectroscopic factors for the $p_{1/2}$ orbit are over predicted while those for the $f_{7/2}$ are too small in the calculation. Overall a good agreement is reached.

TABLE VI: Spectroscopic factors for one neutron transfer from the ground state of $^{182}W$ to various excited states in $^{183}W$ are compared with the data [1].

| $^{183}W$ | Exp. | | Theo. | |
|---|---|---|---|---|
| $j^\pi$ | E | $S_j \times 100$ | E | $S_j \times 100$ |
| $\frac{1}{2}^-_1$ | 0.0 | .55 | 0.0 | 4.2 |
| $\frac{1}{2}^-_2$ | 0.935 | 1.5 | 0.746 | 3.8 |
| $\frac{3}{2}^-_1$ | 0.046 | 8.9 | 0.039 | 6.7 |
| $\frac{3}{2}^-_2$ | 0.209 | 5.8 | 0.213 | 2.1 |
| $\frac{3}{2}^-_3$ | 1.150 | 10. | 0.213 | 43. |
| $\frac{5}{2}^-_1$ | 0.099 | 15 | 0.111 | 8. |
| $\frac{5}{2}^-_2$ | 0.292 | 8.4 | 0.111 | 15. |
| $\frac{7}{2}^-_1$ | 0.207 | – | 0.198 | 0.1 |
| $\frac{7}{2}^-_2$ | 0.412 | 4.4 | 0.395 | 0.2 |
| $\frac{7}{2}^-_3$ | 0.453 | 11.5 | 0.456 | 0.3 |
| $\frac{7}{2}^-_4$ | 1.000 | 2.0 | 0.822 | 0.7 |
| $\frac{13}{2}^+_1$ | 0.486 | 9.7 | 0.426 | 15.8 |

## VII. CONCLUSIONS

We have presented a complete calculation of excitation energies, B(E2) values and spectroscopic factors for neutron stripping in the IBFA model for the odd-mass $^{183-187}W$ isotopes. In general we have obtained a very good agreement of the excitation energies with the experimental data for both positive and negative parity states using a single parametrization for the Hamiltonian. The data for B(E2) values is however scarce and this thus does not offer a very valuable test of the calculations. The recent data on spectroscopic factors offer a nice alternative to test the structure of the wave functions. The calculated spectroscopic factors show some discrepancies in the absolute values for some orbits. The relative magnitudes agree rather well with the data.

**Acknowledgments**

S.A.M. expresses his gratitude to Prof. Daw Saad Mosbah of the Arab Atomic-Energy Agency (AAEA) and thanks him for many useful discussions. We thank P. Van Isacker for a critical reading of the manuscript.

---

[1] V. Bondarenko *et al.*, Nucl. Phys. **A856**, 1 (2011).

[2] V. Bondarenko *et al.*, Nucl. Phys. **A811**, 28 (2008).

[3] T. Shizuma *et al.*, Phys. Rev. C **77**, 047303 (2008).

[4] F. Iachello and P. Van Isacker, (Cambridge Univ., New York, 1991).

[5] F. Iachello and A. Arima, 'The Interacting Boson Model', (Cambridge Univ. Press, 1987)

[6] P. D. Duval and B. R. Barrett, Phys. Rev. C **23**, 492 (1981).

[7] L. Esser *et al.*, Phys. Rev. C **55**, 206 (1997).

[8] C. Wheldon *et al.*, Phys. Rev. C **63**, 011304 (2001).

[9] C. Thwaites *et al.*, Phys. Rev. C **66**, 054309 (2002).

[10] E. A. McCutchan and N. V. Zamfir, Phys. Rev. C **71**, 054306 (2005).

[11] E. Ngijoi-Yogo *et al.*, Phys. Rev. C **75**, 034305 (2007).

[12] U. Kaup, A. Gelberg, P. von Brentano and O. Scholten, Phys. Rev. C **22** (1980) 1738.

[13] O. Scholten and N. Blasi, Nucl. Phys. **A380**, 509 (1982). O. Scholten and T. Ozzello, Nucl. Phys. **A424**, 221 (1984).

[14] J. Jolie, *et al.*, Nucl. Phys. **A438**, 15 (1985).

[15] B. Warner *et al.*, Phys. Rev. Lett. **54**, 1365 (1985).

[16] I. Shestakova *et al.*, Phys. Rev. C **64**, 054307 (2001).

[17] W. Nazarewicz, M. A. Riley, and J. D. Garrett, Nucl. Phys. **A512**, 61 (1990).

[18] F. Iachello and O. Scholten, Phys. Rev. Lett. **43** (1979) 679.

[19] U. Kaup *et al.*, Phys. Lett. B **106**, 439 (1981).

[20] O. Scholten and S. Pittel, Phys. Lett. B **120**, 9 (1983).

[21] O. Scholten, Prog. in part. and nuc. phys., ed. A Faessler (1985) vol. 14, p. 189

[22] A. E. L. Dieperink, O. Scholten and F. Iachello, Phys. Rev. Lett. **44** (1980) 1747.

[23] R. Bijker and O. Scholten, Phys. Rev. C **32** (1985) 591.

[24] F. Iachello and O. Scholten, Phys. Lett. B **91**, 189 (1980).

[25] O. Scholten, Phys. Lett. B **108**, 155 (1982).

[26] P. Van Isacker, *et al.*, Phys. Lett. B **149**, 26 (1984).

[27] W. T. Chou *et al.*, Phys. Rev. C **42** (1990) 221.

[28] P.F. Bortignon, G. Colò, and H. Sagawa, J. Phys. G**37**, 064013 (2010).

[29] H.R. Yazar, J. Korean Phys. Soc. 49(1) , 60 (2006)

[30] http://ie.lbl.gov/TOI2003/GammaSearch.asp

[31] B. S. Reehal and R. A. Sorensen, Phys. Rev. C2, 819 (1970)

[32] O. Scholten, Computer code PHINT, KVI (Groningen, The Netherlands, 1980)

[33] http://www.nndc.bnl.gov/

[34] I. Talmi, Interacting Bose-Fermi system in nuclei, ed. F. Iachello (Plenum Press, New York, 1981), p. 329

[35] O. Scholten, Internal Report KVI 252 Computer Code ODDA, University of Groningen, 1980.